\documentclass[%
 reprint,
superscriptaddress,
showpacs,preprintnumbers,
 amsmath,amssymb,
 aps,
pra,
]{revtex4-1}

\usepackage{xcolor}
\usepackage{soul}
\usepackage{graphicx}  
\usepackage{dcolumn}  
\usepackage{bm}  

\newcommand{\sr}{\mathrm{s}}			
\newcommand{\pr}{\mathrm{p}}			
\newcommand{\dr}{\mathrm{d}}			
\newcommand{\SR}{\mathrm{S}}			
\newcommand{\DR}{\mathrm{D}}			
\newcommand{\ket}[1]{| \, #1\, \rangle}							
\newcommand{\matel}[3]{\langle \, #1 \, | \, #2 \, | \, #3 \, \rangle}          
\newcommand{\braket}[2]{\langle \, #1 \, | \, #2 \, \rangle}		
\newcommand{\ex}{\hat{e}_x}			
\newcommand{\ey}{\hat{e}_y}			
\newcommand{\ez}{\hat{e}_z}			
\DeclareMathAlphabet\mathbfcal{OMS}{cmsy}{b}{n}         
\newcommand{\ver}{\mathbf{r}}			
\newcommand{\cc}{\mathrm{c.c.}}         
\newcommand{\er}{\mathrm{e}}			
\newcommand{\ir}{\mathrm{i}}			
\newcommand{\ra}[1]{\renewcommand{\arraystretch}{#1}}           
\renewcommand{\Re}{\operatorname{Re}}

\renewcommand{\vec}{\bm}

\begin{document}


\title{An optical lattice based method for precise measurements of atomic parity violation}

\author{A. Kastberg}
\email{anders.kastberg@unice.fr}
\affiliation{%
Institute de Physique de Nice, Universit\'e C\^ote d'Azur, CNRS, 06108 Nice, France
}%

\author{T. Aoki}%
\affiliation{
Institute of Physics, Graduate School of Arts and Sciences, The University of Tokyo, Tokyo, Japan
}%

\author{B. K. Sahoo}
\affiliation{ Atomic, Molecular and Optical Physics Division, 
Physical Research Laboratory, Navrangpura, Ahmedabad 380009, India
}%

\author{Y. Sakemi}
\affiliation{%
Center for Nuclear Study, The University of Tokyo, Wako, Japan
}%

\author{B. P. Das}
\affiliation{%
International Education and Research Center of Science and Department of Physics, Tokyo Institute of Technology, Tokyo, Japan
}%


\date{\today}

\begin{abstract}
We propose a method for measuring parity violation in neutral atoms. It is an adaptation of a seminal work by Fortson [Phys. Rev. Lett. {\bf 70}, 2383 (1993)], proposing a scheme for a single trapped ion. In our version, a large sample of neutral atoms should be localised in an optical lattice overlapping a grid of detection sites, all tailored as the single site in Fortson's work. The methodology is of general applicability, but as an example, we estimate the achievable signal in an experiment probing a nuclear spin independent parity violation on the line $6\sr\,^2\SR_{1/2}$--$5\dr\,^2\DR_{3/2}$ in $^{133}$Cs. The projected result is based on realistic parameters and \textit{ab initio} calculations of transition amplitudes, using the relativistic coupled-cluster method. The final result is a predicted spectroscopic signature, evidencing parity violation, of the order of 1 Hz, for a sample of $10^8$ atoms. We show that a total interrogation time of 30000 s should, together with existing theoretical data, suffice for achieving a precision in the determination of the weak charge of Cs of the order of 0.1\% --- a sensitivity surpassing previously reported determinations by at least a factor of five. 
\end{abstract}



\maketitle


Parity non-conserving (PNC) interactions in atomic systems involve interplay between the weak and electromagnetic forces \cite{Bouchiat1997}, and its studies have implications for atomic, nuclear and particle physics \cite{Ginges2004,Safronova2018}. They enable explorations of new physics beyond the Standard Model (SM) of particle physics \cite{Commins2014,Erler2010}, probe the nuclear anapole moment (NAM) --- presumed to be a fundamental property of a nucleus \cite{Ginges2004,Zeldovich1957,Flambaum1980} --- and test the role of electron correlation effects in atomic systems \cite{Safronova2018}. For detection of PNC signatures in atomic transitions, the underlying phenomena can be categorised as nuclear spin dependent (NSD) or independent (NSI), providing means to tailor experiments for specific purposes. Underlying causes for parity violations are the exchange of a $Z_0$ boson between an electron and the nucleus, due to the weak interaction, and the NAM. The $Z_0$-current has both NSI and NSD contributions, whereas the spectral effects of the NAM are purely NSD in character \cite{Ginges2004}.

When choosing a system for an NSI PNC experiment, an important consideration is that the PNC interaction scales approximately as the third power of the nuclear charge \cite{Bouchiat1974}. As a consequence, a heavy atom will greatly facilitate detection. Secondly, extraction of the weak charge from PNC signatures requires knowledge of the NSI transition amplitude and thus it has advantages to choose a system amenable to accurate computational efforts. These considerations make a heavy alkali system a good choice. One reported study --- the only claiming observation of a NAM --- used Cs and the dipole forbidden transition $6\sr\,^2\SR_{1/2}$--$7\sr\,^2\SR_{1/2}$ \cite{Wood1997}. However, the observed NAM 
is at variance with predictions based on the shell model and nucleon-nucleon scattering experiments \cite{Wilburn1998,Haxton2001}, motivating a revisit of the subject. In another reported PNC experiment the transition $6\sr^2\,^1\SR_0$--$5\dr6\sr\,^3\DR_1$ in Yb is used \cite{Tsigutkin2009}. 

The PNC interaction induces mixing of states, resulting in non-zero matrix elements between levels otherwise lacking a dipole allowed transition. This transition amplitude is not suitable for direct detection, and many works focus instead on an interference between this and another excitation amplitude. In \cite{Wood1997} and \cite{Tsigutkin2009}, this pertained to Stark induced magnetic dipole (M1) resonances. Here we analyse a PNC amplitude interfering with an electrical quadrupole transition (E2) on a $\sr$--$\dr$ spectral line in an alkali, and we complement a concrete experimental scheme by showing that for Cs, the transition 6s--5d holds advantages for PNC experiments over 6s--7s \cite{Dzuba2001}. Attempts to detect PNC amplitudes, using $\sr$--$\dr$ transitions in alkali systems have focused on ions. In a seminal work \cite{Fortson1993}, Fortson introduced a scheme with a single trapped Ba$^+$ ion on the line $6\sr\,^2\SR_{1/2}$--$5\dr\,^2\DR_{3/2}$. The choice of a single ion limits the statistics, but this is partly offset by the long lived upper state and the long storage time. Here, we adapt the scheme in \cite{Fortson1993} to neutral atoms trapped in an optical lattice, thereby enabling a substantial reduction of shot-noise limitations. We will trap individual atoms periodically, and with auxiliary fields, independent of the trapping light, we will follow the detection idea of \cite{Fortson1993}. In order to demonstrate the feasibility of the suggestion, we analyse a concrete example of a measurement of NSI in Cs, including calculation of $6\sr\,^2\SR_{1/2}$--$5\dr\,^2\DR_{3/2}$ transition amplitude.

\begin{figure}
\includegraphics[width=0.70\columnwidth]{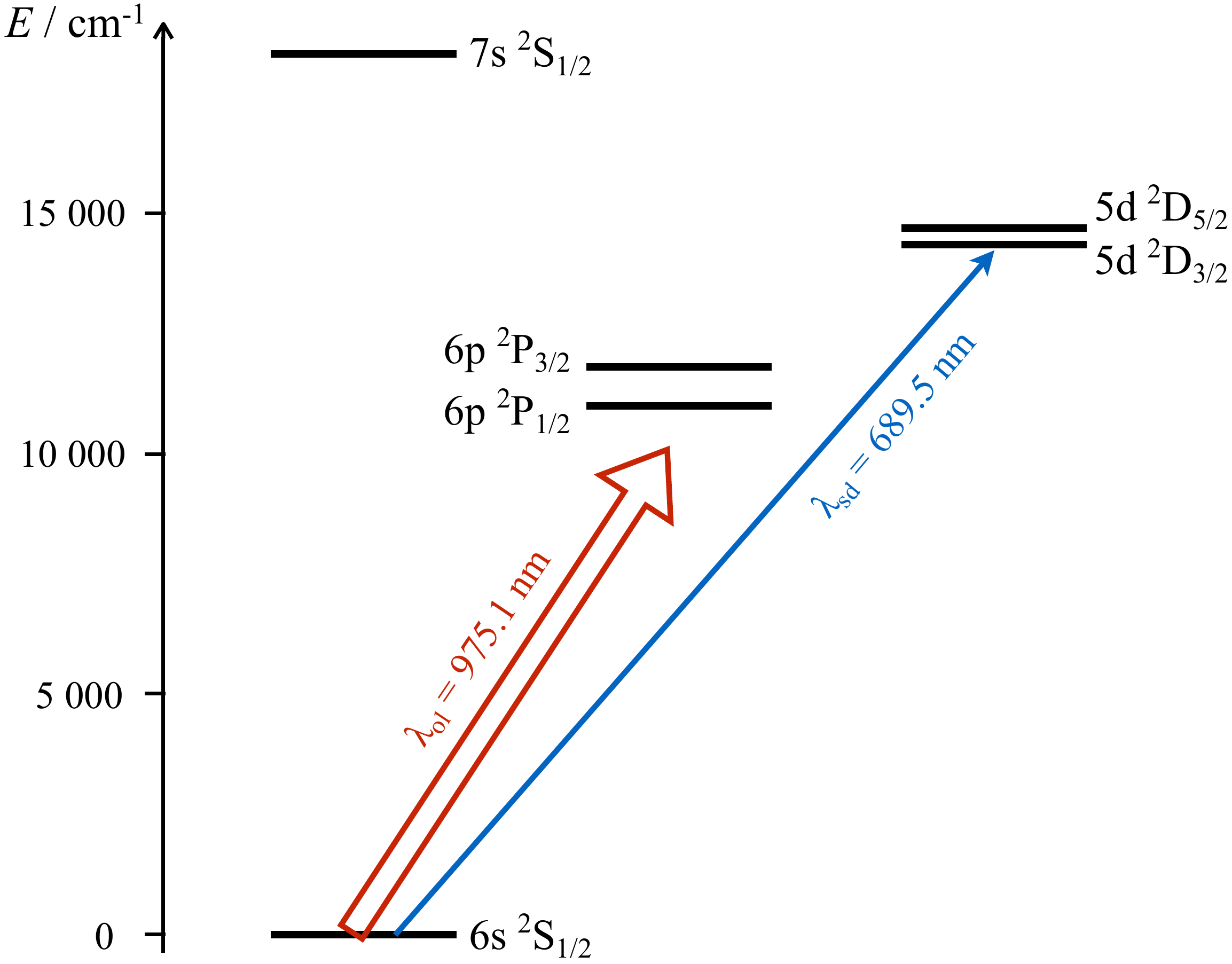}
\caption{Energy levels corresponding to the lowest configurations (6s, 6p, 5d, and 7s) in Cs I, including fine-structure. The PNC detection will be performed with two lasers tuned to the resonance $6\sr\,^2\SR_{1/2}$--$5\dr\,^2\DR_{3/2}$ at $\lambda_\mathrm{sd}\!=\!689.5$ nm. The optical lattice must operate on the wavelength $\lambda_\mathrm{ol}\!=\!\lambda_\mathrm{sd}\sqrt{2}=975.1$ nm to achieve overlap with points suitable for detection.}
\label{fig_Cs}
\end{figure}

We suggest trapping atoms in a two-dimensional optical lattice, where each trapping site must coincide with a detection field tailored to optimise a PNC signature \cite{Fortson1993}. The latter criterion is that a nodal plane from one standing wave crosses an anti-nodal one from a second one with the same wavelength, resonant with $6\sr\,^2\SR_{1/2}$--$5\dr\,^2\DR_{3/2}$ ($\lambda_\mathrm{sd}\!=\!689.5$ nm) --- see Grotrian diagram in Fig.\ \ref{fig_Cs}. 
This takes advantage of the fact that the oscillator strength of an E2-transition is proportional to the square of the electric field gradient, while that of the PNC induced E1-transition scales linearly with intensity. For both, the standing wave configurations enhance the excitation rates and associated light shifts at the sites optimised for detection. The relevant Rabi frequencies for the E2 and PNC excitations are:
\begin{align}
	\Omega_\mathrm{E2} &= -\frac{1}{2\hbar} \sum_{i,j} (A_\mathrm{E2})_{i,j} \, \left[\frac{\partial\mathcal{E}_i(\ver)}{\partial x_j}\right]_{\ver=\vec{0}}  \notag \\
    \mathrm{and} \quad \Omega_\mathrm{PNC} &= -\frac{1}{2\hbar} \sum_i (A_\mathrm{PNC})_i \; \mathcal{E}_i(\ver\!=\!\vec{0})  \; .
\end{align}
The indices represent the Cartesian coordinates and the tensor elements $(A_\mathrm{E2})_{i,j}$ and $(A_\mathrm{PNC})_i$ are the E2 and PNC amplitudes. The origin is chosen to be at the centre of one of the good detection volumes, and $\mathcal{E}_i$ is the electric field component along $i$. With the transitions simultaneously driven, and provided that the driving field is resonant, the overall light shifts are \cite{Aoki2017}:
\begin{align}
\label{eq_LightShiftE2PNC}
	\Delta E &= \hbar \,\sqrt{| \Omega_\mathrm{E2} + \Omega_\mathrm{PNC} |^2}  \notag \\
	&\approx \hbar \,\Omega_\mathrm{E2} + \frac{\hbar\,\Re\left[\Omega_\mathrm{E2}\Omega^*_\mathrm{PNC}\right]}{\Omega_\mathrm{E2}} \equiv  W_\mathrm{E2} + W_\mathrm{PNC}   \; .
\end{align}
Here, the pure PNC term has been neglected, as has contributions from M1 amplitudes.

\begin{figure*}
\includegraphics[width=1.9\columnwidth]{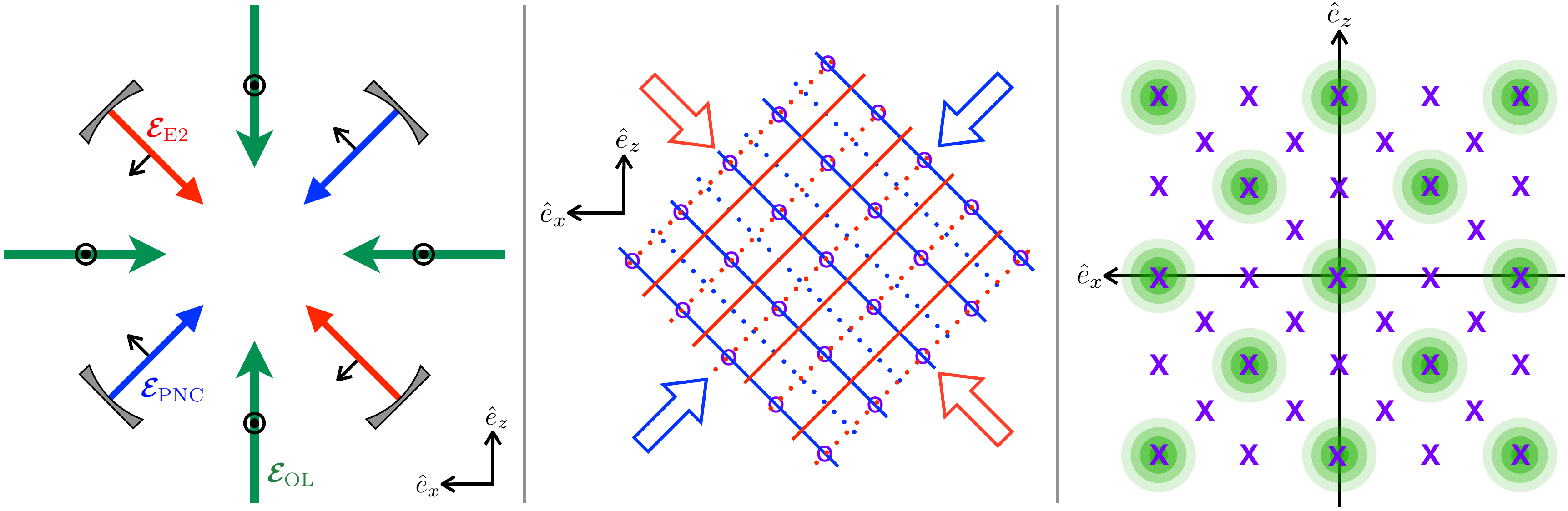}
\caption{Left: Proposed laser configurations. Red and blue arrows are cavity wave beam pairs, at angles of $\pi/4$ with the $\ex$ and $\ez$ axes, which drive the E2 and PNC amplitudes. These fields have mutually orthogonal linear polarisations, and their common wavelength $\lambda_\mathrm{sd}$ is resonant with the 6s--5d transition. Green arrows represent optical lattice beams, linearly polarised along $\ey$ and with wavelength $\lambda_\mathrm{ol}\!=\!\lambda_\mathrm{sd}\sqrt{2}$. Centre: Schematic illustration of two overlapping orthogonal standing waves, resulting in points that are optimised for detection of PNC. The PNC-field is shown in blue and the E2-field in red. For both, anti-nodal planes are shown as full lines and nodal points as dotted lines. The sites optimised for detection are ones where a PNC anti-node coincide with an E2 node (indicated by purple circles. Right: Locations in the $xz$-plane of points optimised for PNC detection (purple `X'-symbols) and optical lattice sites (green concentric, filled circles). For the optical lattice, darker green represent higher irradiance.}
\label{fig_setup}
\end{figure*}

A configuration that fulfils the criteria above is shown in Fig.\ \ref{fig_setup}.
The laser fields that drive the E2 and PNC excitation amplitudes are, with coordinates and orientations as in the figure: 
\begin{align}
\label{eq_E2PNCfield}
	\mathbfcal{E}_\mathrm{E2}(\ver,t) &= \frac{\mathcal{E}_\mathrm{E2}}{2\sqrt{2}} \left(\ex\!-\!\ez\right) \left\{ \exp\!\left[ -\frac{\ir k_\mathrm{sd}}{\sqrt{2}} \left(\ex\!+\!\ez\right)\!\cdot\!\ver \right] \right.  \notag \\
    + &\left. \exp\!\left[\frac{\ir k_\mathrm{sd}}{\sqrt{2}} \left(\ex\!+\!\ez\right)\!\cdot\!\ver + \ir\pi\right] \right\} \er^{\ir\omega_\mathrm{sd} t} + \cc  \notag \\
    \mathbfcal{E}_\mathrm{PNC}(\ver,t) &= \frac{\mathcal{E}_\mathrm{PNC}}{2\sqrt{2}} \left(\ex\!+\!\ez\right) \left\{ \exp\!\left[ -\frac{\ir k_\mathrm{sd}}{\sqrt{2}} \,\left(\ex\!-\!\ez\right)\!\cdot\!\ver \right] \right.  \notag \\
    + &\left. \exp\!\left[ \frac{\ir k}{\sqrt{2}} \left(\ex\!-\!\ez\right)\!\cdot\!\ver \right] \right\} \er^{\ir\omega_\mathrm{sd} t} + \cc  \enspace .
\end{align}
$\mathcal{E}_\mathrm{E2}$ and $\mathcal{E}_\mathrm{PNC}$ are the electric field amplitudes, $\omega_\mathrm{sd}$ the angular frequency, and $k_\mathrm{sd}\!=\!\omega_\mathrm{sd}/c$ the angular wave number. The relative phase of $\pi$ ensures that a nodal plane crosses the origin. 
The light in Eq.\ (\ref{eq_E2PNCfield}) results in a two-dimensional lattice of points amenable for PNC detection, as illustrated in Fig.\ \ref{fig_setup}.

The role of the optical lattice is to keep every atom localised in the $xz$-plane, much tighter than $\lambda_\mathrm{sd}^{\,2}$, centred at one of the points tailored for PNC detection. Our proposed configuration is four laser beams oriented along the $\ex$ and $\ez$ axes:
\begin{multline}
\label{eq_OLfield}
	\mathbfcal{E}_\mathrm{ol}(\ver,t) = \frac{\mathcal{E}_\mathrm{ol}}{2} \,\ey \left\{ \exp\!\left[ - \ir k_\mathrm{ol}\,\ex\!\cdot\!\ver\,\right] + \exp\!\left[ \,\ir k_\mathrm{ol}\,\ex\!\cdot\!\ver\,\right] \right. \\
	\left. + \exp\!\left[ -\ir k_\mathrm{ol}\,\ez\!\cdot\!\ver\,\right] + \exp\!\left[\, \ir k_\mathrm{ol}\,\ez\!\cdot\!\ver\,\right] \right\} \er^{\ir\omega_\mathrm{ol}t} + \cc  \enspace .
\end{multline}
The angular frequency needed to have the trapping sites coincide with the points suitable for detection, defined in Eq.\ (\ref{eq_LightShiftE2PNC}), is $\omega_\mathrm{ol}\!=\!\omega_\mathrm{sd}/\sqrt{2}$, and $\mathcal{E}_\mathrm{ol}$ is the amplitude per beam. The temporal phases will be controlled interferometrically \cite{Hemmerich1993}. This leads to a light shift potential \cite{Hemmerich1993,Grynberg2001,Greiner2001}:
\begin{multline}
\label{eq_OL}
	U(\ver) = U_0 \left[ \cos^2(k_\mathrm{ol}x) + \cos^2(k_\mathrm{ol}z)  \right. \\
    \left. + 2 \cos(k_\mathrm{ol}x) \cos(k_\mathrm{ol}z) \right]  \; ,
\end{multline}
where $U_0$ is the light shift at irradiance maxima. With the chosen frequency, the optical lattice, at $\lambda_\mathrm{ol}\!=\!\lambda_\mathrm{sd}\sqrt{2}=975.1$ nm, is detuned below the principal E1-resonances, 6s--6p. 
Thus, the light shift will be negative, and the potential minima will be at irradiance maxima. Laser cooled atoms will now be confined around points in the $xz$-plane that are commensurate with the good detection points as illustrated in Fig.\ \ref{fig_setup}. There will be many optimised detection points lacking a trapped atom, but it is only the reverse requirement which is a necessary condition. The chosen geometry carries the extra advantages that it enables cancellations of contributions to the measured PNC amplitude from the light shift induced by the optical lattice, and from the first order Zeeman shift. This will be detailed in the forthcoming.

The temporal phases of all laser beams in Eqs.\ (\ref{eq_E2PNCfield}) and (\ref{eq_OLfield}) will be controlled interferometrically by electronic feedback to all end mirrors of the beams in Fig.\ \ref{fig_setup}. The two lasers running at $\lambda_\mathrm{sd}$ and $\lambda_\mathrm{ol}$ do not have to be phase locked, since a global phase drift in any of them will appear equally in all interferometer arms. That means that all phase drifts before the respective beams are split will cancel (see \cite{Ellman2003} or \cite{Sjolund2006}).  

\begin{figure}
\includegraphics[width=0.95\columnwidth]{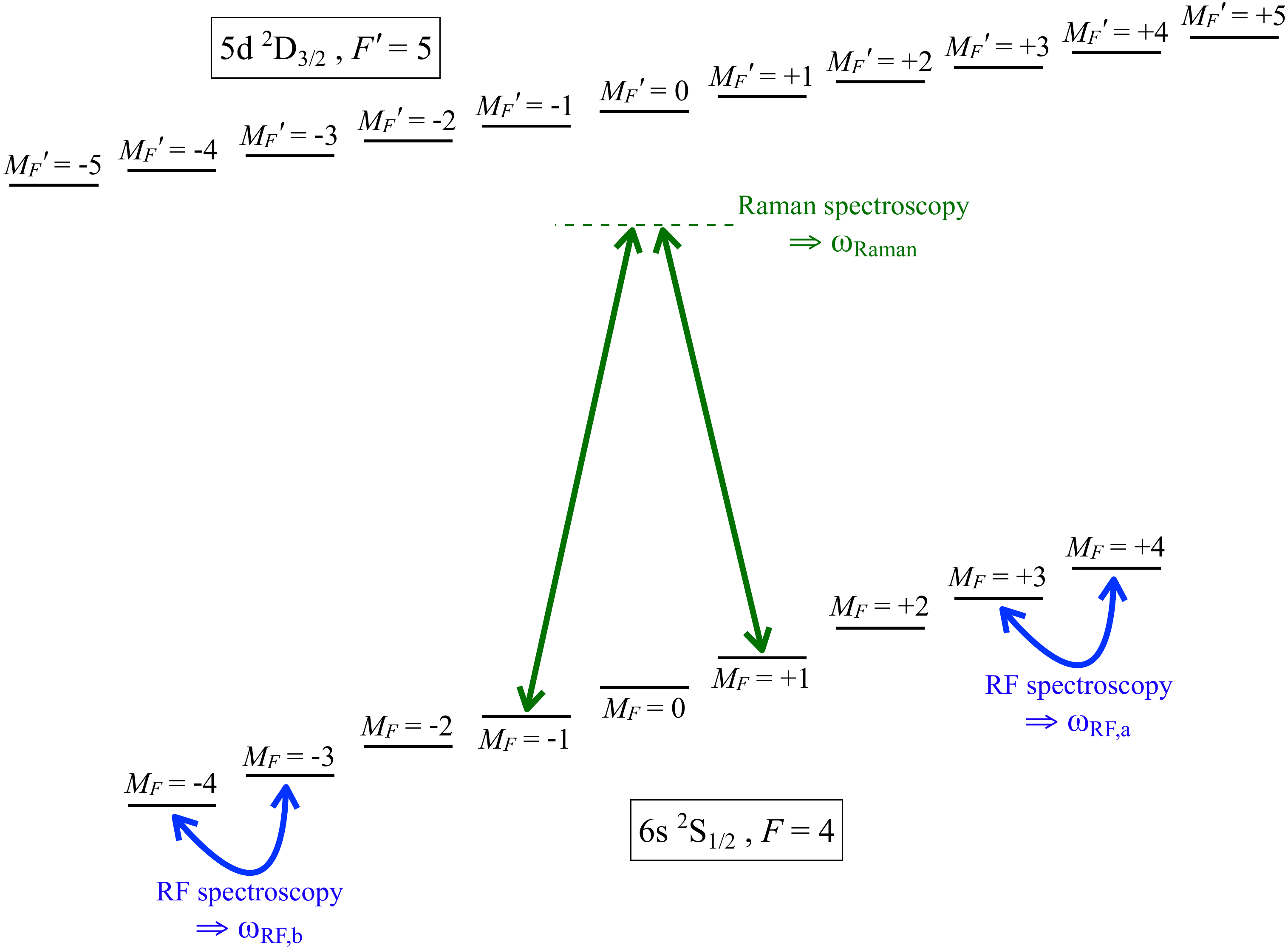}
\caption{Illustration of the suggested detection configuration. The total induced level separation is measured between three pairs of magnetic sublevles in $6\sr\,^2\SR_{1/2},F\!=\!4$: $\hbar\,\omega_\mathrm{RF,a} \!=\! \Delta E(+4) \!-\! \Delta E(+3)$ and $\hbar\,\omega_\mathrm{RF,b} \!=\! \Delta E(-4) \!-\! \Delta E(-3)$ by RF-spectroscopy, and $\hbar\,\omega_\mathrm{Raman} \!=\! \Delta E(+1) \!-\! \Delta E(-1)$ by Raman spectroscopy.}
\label{fig_SUBevels}
\end{figure}

As a concrete example, supporting the feasibility of the suggestion, we consider a measurement of the NSI light shifts of the magnetic substates $M_F\!=\!\pm1$, $\pm3$ and $\pm4$ of the Cs level $6\sr\,^2\SR_{1/2},F\!=\!4$, as illustrated in Fig.\ \ref{fig_SUBevels}. 
These should be brought into resonance with states of the excited level $5\dr\,^2\DR_{3/2},F^\prime\!=\!5$, using lasers at $\lambda_\mathrm{sd}\!=\!689.5$ nm. All included sublevels will be shifted from their field-free energies due to light shifts induced by the E1-interaction with the optical lattice light, and the PNC E1-coupling and E2-interaction induced by the driving fields. Added to that will be Zeeman shifts, induced by a weak magnetic field that establishes the quantisation axis. Provided zero detunings of E2 and PNC lasers, the total level dependent energy shifts, adjusted from Eq.\ (\ref{eq_LightShiftE2PNC}) will be:
\begin{multline}
\label{eq_DeltaEtotal}
	\Delta E(M_F) = M_F E_\mathrm{Z} + E_{qZ,M}  \\
	+ U_{0,M} + W_\mathrm{E2,M} + W_\mathrm{PNC,M}   \; ,
\end{multline}
wherein $E_\mathrm{Z}$ is the Larmor frequency in energy  units; and $E_{qZ\!,M}$, $U_{0,M}$, $W_\mathrm{E2,M}$ and $W_\mathrm{PNC,M}$  respectively the quadratic Zeeman shift, optical lattice potential and the E2 and PNC light shifts for a specific Zeeman sublevel. For the nine different Zeeman levels individually, the light shifts are:
\begin{align}
\label{eq_ZeemanLevelShifts}
    \Delta E(+4) &= 4\,E_\mathrm{Z} + U_{0,4} + W_\mathrm{E2,3} + W_\mathrm{PNC,4}  \notag \\
	\Delta E(+3) &= 3\,E_\mathrm{Z} + E_{qZ,3} + U_{0,3} + W_\mathrm{E2,3} + W_\mathrm{PNC,3}  \notag \\
	\Delta E(+2) &= 2\,E_\mathrm{Z} + E_{qZ,2} + U_{0,2} + W_\mathrm{E2,2} + W_\mathrm{PNC,2}  \notag \\
	\Delta E(+1) &= \;E_\mathrm{Z} + E_{qZ,1} + \,U_{0,1} + \,W_\mathrm{E2,1} + \,W_\mathrm{PNC,1}  \notag \\
	\Delta E(0)\; &= \;E_{qZ,0} + \,U_{0,0} + \,W_\mathrm{E2,0} + \,W_\mathrm{PNC,0}  \notag \\	
	\Delta E(-1) &= -E_\mathrm{Z} + E_{qZ,-1} + U_{0,1} + W_\mathrm{E2,1} - W_\mathrm{PNC,1}  \notag \\
	\Delta E(-2) &= -2\,E_\mathrm{Z} + E_{qZ,-2} + U_{0,2} + W_\mathrm{E2,2} - W_\mathrm{PNC,2}  \notag \\	
	\Delta E(-3) &= -3\,E_\mathrm{Z} + E_{qZ,-3} + U_{0,3} + W_\mathrm{E2,3} - W_\mathrm{PNC,3}  \notag \\    
	\Delta E(-4) &= -4\,E_\mathrm{Z} + U_{0,4} + W_\mathrm{E2,3} - W_\mathrm{PNC,4}  \enspace .
\end{align}

We consistently use $\ez$ as quantisation axis. The optical lattice field can then be decomposed in $\sigma_+$ and $\sigma_-$ fields, driving only $\Delta M_F\!=\!\pm1$ light shifts. This means that $U_{0,M}\!=\!U_{0,-M}$ \cite{Chin2001}. For the E2-transitions, with geometry and polarisations as in Fig.\ \ref{fig_setup},  $\Delta M_F\!=\!0$ will dominate \cite{Aoki2017}. $\Delta M_F\!=\!\pm2$ will contribute little, and $\Delta M_F\!=\!\pm1$ will be totally suppressed. With the added magnetic field the contributions from $\Delta M_F\!=\!\pm2$ may be further reduced for a laser tuned to $\Delta M_F\!=\!0$. The E2 interaction is parity conserving, which means that $W_\mathrm{E2,M}\!=\!W_\mathrm{E2,-M}$. Furthermore, the Clebsch-Gordan coefficients are such that $W_\mathrm{E2,\pm3}\!=\!W_\mathrm{E2,\pm4}$. The selection rule for M1 is $\Delta M_F\!=\!\pm1$. With our parameters, and the insignificant M1 amplitude (shown later), M1 contributions can be neglected.

The sought after quantities are $W_\mathrm{PNC,M}$, and since the PNC interaction is parity violating $W_\mathrm{E2,M}\!=\!-W_\mathrm{E2,-M}$. This can be utilised in order to eliminate as many as possible of the technical contributions to Eq.\ (\ref{eq_DeltaEtotal}) and isolate the PNC signature. We propose to measure the splittings between the $M_F\!=\!\pm4$ and $M_F\!=\!\pm3$ with Ramsey RF-spectroscopy:
\begin{align}
\label{eq_OmegaRF}
	\hbar\,\omega_\mathrm{RF,a} &= \Delta E(+4) - \Delta E(+3)  \notag \\
    \mathrm{and} \quad 	\hbar\,\omega_\mathrm{RF,b} &= \Delta E(-4) - \Delta E(-3)  \; ,
\end{align}
and the energy difference between $M_F\!=\!1$ and $M_F\!=\!-1$ with Raman spectroscopy, see Fig.\ \ref{fig_SUBevels}:
\begin{equation}
\label{eq_OmegaRaman}
	\hbar\,\omega_\mathrm{Raman} = \Delta E(+1) - \Delta E(-1)  \; .
\end{equation}
In the latter case, the light shift induced by the Raman beams will cancel if the two beams have the same intensity. The duration of the Ramsey interrogation periods will be set by the radiative lifetime of the upper level, and state selective detection will be done using 9.2 GHz microwave radiation and induced fluorescence (see \cite{Aoki2017} for a detailed description). The three spectroscopical results in Eqs.\ (\ref{eq_OmegaRF}) and (\ref{eq_OmegaRaman}) can then be combined as follows:
\begin{align}
\label{eq_OmegaObs}
	\hbar &\left(\omega_\mathrm{RF,a} - \omega_\mathrm{RF,b} - \omega_\mathrm{Raman}\right)  \notag \\
    &= \left( W_\mathrm{PNC,4} - W_\mathrm{PNC,3} - W_\mathrm{PNC,1} \right) + E_{qZ}  \notag \\
    &= \hbar\omega_\mathrm{obs} + E_{qZ}  \; .
\end{align}
The last term is the total contribution from the quadratic Zeeman shift: 
\begin{equation}
\label{eq_quadraticZeeman}
    E_{qZ} = E_{qZ,-3}+E_{qZ,-1}-E_{qZ,1}-E_{qZ,3}  \; .
\end{equation}
It cancels for $M_F\!=\!\pm4$, and for the remaining levels, the contribution to the signal can be accurately determined by performing the same spectroscopy as above, but without the E2 and PNC driving fields. The remaining energy, $\hbar\omega_\mathrm{obs}$, provides a measurement of the PNC transition amplitude, $A_\mathrm{PNC}$. 

In order to estimate the achievable signal, we have calculated $A_\mathrm{PNC}$, $A_\mathrm{E2}$, and the amplitude due to M1 ($A_\mathrm{M1}$) of the $6\sr\,^2\SR_{1/2}$--$5\dr\,^2\DR_{3/2}$ line, using a relativistic coupled-cluster (RCC) theory. Similar results were previously reported using a sum-over-states approach \cite{Sahoo2017}. The latter calculations are here improved by solving the first-order perturbed equations in the presence of PNC. In the RCC theory, we express the initial and final states of the transition, without considering the PNC interaction, as: 
\begin{equation}
\label{eq_eqcc}
    \ket{\Psi_v^{(0)}} = \er^{T^{(0)}} [ 1+ S_v^{(0)} ] \,\ket{\Phi_v}  \; .
\end{equation}
Here $\ket{\!\Phi_v\!}\!=\!a_v^{\dagger} \ket{\!\Phi_0\!}$ is the Dirac-Hartree-Fock (DHF) wave function of the $5\pr^6$ closed-shell configuration, and $v$ corresponds to the respective valence orbitals of the initial and final states. The RCC operators, $T^{(0)}$ and $S_v^{(0)}$ excite electrons from $\ket{\!\Phi_0\!}$ and $\ket{\!\Phi_v\!}$, respectively, to the virtual space. The M1 and E2 matrix elements are determined in a similar approach as in \cite{Sahoo2017}, but the accuracies are improved.

The $A_\mathrm{PNC}$ amplitude between states with valence orbitals i and f can be evaluated as:
\begin{equation}
    A_\mathrm{PNC}= \frac{ \matel{\Psi_\mathrm{f}^{(0)} \!+\! \Psi_\mathrm{f}^{(1)}}D{\Psi_\mathrm{i}^{(0)} \!+\!  \Psi_\mathrm{i}^{(1)}} }{ \sqrt{\braket{ \Psi_\mathrm{f}^{(0)}}{\Psi_\mathrm{f}^{(0)}} \braket{\Psi_\mathrm{i}^{(0)}}{\Psi_\mathrm{i}^{(0)}}} }  \; ,
\end{equation}
where $D$ is the E1 operator, and superscript $1$ denotes the first-order PNC perturbed wave functions with respect to the DC Hamiltonian. In the sum-over-states approach used in \cite{Sahoo2017}, the latter state is expressed as:
\begin{equation}
    \ket{ \Psi_v^{(1)} } = \sum_{k\neq v} \ket{ \Psi_k^{(0)} } \: \frac{ \matel{\Psi_k^{(0)}}{H_\mathrm{PNC}^\mathrm{NSI}}{\Psi_v^{(0)}} }{ E_v^{(0)}-E_k^{(0)} }  \; ,
\end{equation}
where $H_\mathrm{PNC}^\mathrm{NSI}$ represents the NSI PNC interaction Hamiltonian. The sum is over all allowed intermediate eigenstates of the atomic Hamiltonian $H_\mathrm{a}$, with energies $E_{k=\text{i,f}}^{(0)}$. This restricts the sum to matrix elements containing only low-lying bound states in the RCC theory, while contributions from the continuum are included using lower-order many-body methods. To circumvent this, we have instead obtained first-order perturbed wave functions by solving an inhomogeneous equation for a  
state involving the valence orbital $v$: 
\begin{equation}
    (H_\mathrm{a} - E_v^{(0)}) \,\ket{ \Psi_v^{(1)} } = (E_v^{(1)} - H_\mathrm{PNC}^\mathrm{NSI}) \,\ket{ \Psi_v^{(0)} }  \: .
\end{equation}
The first-order energy perturbation, $E_v^{(1)}$, vanishes due to the odd-parity of $H_\mathrm{PNC}^\mathrm{NSI}$. In the RCC theory framework, this result is obtained by: 
\begin{equation}
 \label{eq_eqrcc}
    \ket{ \Psi_v^{(1)} }  = \er^{T^{(0)}} \{ T^{(1)} \,[1 + S_v^{(0)}] + S_v^{(1)}\} \, \ket{ \Phi_v }  \; ,
\end{equation}
where superscript $1$ now means the first-order perturbed RCC operators with respect to $H_\mathrm{PNC}^\mathrm{NSI}$. The amplitude equations both for the unperturbed and perturbed RCC operators are given in \cite{Sahoo2006,Wansbeek2008}.

We use several methods (in order to check the consistency): DHF; RCC with a single and double excitations approximation (RCCSD); RCC with single, double and triple excitations (through $S_v$) approximation (RCCSDvT); and the approximation with all possible single, double and triple excitations  (RCCSDT), employing the Dirac-Coulomb (DC) Hamiltonian and single particle orbitals generated by Gaussian type orbitals (GTOs) \cite{Sahoo2017}. We have estimated corrections from the Breit interaction, nuclear structure, and lower-order quantum electrodynamics (QED) effects due to the vacuum polarisation and self-energy interactions, using the expressions given in \cite{Sahoo2016B,Ginges2016} in the RCCSD method. Results from different methods and relativistic corrections to $A_\mathrm{PNC}$, $A_\mathrm{M1}$ and $A_\mathrm{E2}$ amplitudes are given in Table \ref{tab_transamp} along with approximate of errors. These uncertainties are determined by analysing errors stemming from the use of finite size GTOs and estimating contributions from neglected higher level excitations. 
\begin{table}
\begin{center}
\caption{Contributions to the overall $6\sr \,^2\SR_{1/2}$--$5\dr \,^2\DR_{3/2}$ transition amplitude from the M1, E2, and PNC interactions, with the DC Hamiltonian derived by different methods. The M1 and E2 amplitudes are in a.u., and the PNC one in $-\ir ea_{0} [Q_\mathrm{W}/N]\!\times\!10^{-11}$ ($Q_\mathrm{W}$: weak charge, $N$: neutron number).
Corrections from Breit and QED interactions are computed with the RCCSD method and nuclear structure corrections are estimated by varying the Fermi nuclear charge charge distribution.}
\label{tab_transamp}
\ra{1.1}
\begin{tabular}{l c c c c c c}
\hline 
Method  &\phantom{a}&  \multicolumn{1}{c}{$A_\mathrm{M1}$}  &\phantom{a}&  $A_\mathrm{E2}$  &\phantom{a}&  $A_\mathrm{PNC}$ \\
\hline
\multicolumn{7}{c}{DC Hamiltonian} \\
DHF  &&  $\sim 0$  &&  43.85  &&  2.376 \\
RCCSD  &&  $2.56 \!\times\! 10^{-4}$  &&  33.98  &&  3.169 \\
RCCSDvT  &&  $2.59 \!\times\! 10^{-4}$  &&  33.94  &&  3.163 \\
RCCSDT   &&  $2.80 \!\times\! 10^{-4}$  &&  33.89  &&  3.165 \\
\hline 
\multicolumn{7}{c}{Corrections} \\
Breit  &&  $7.0 \!\times\! 10^{-5}$  &&  $-0.04$  &&  $-0.017$\\
QED  &&  $-3.0 \!\times\! 10^{-5}$  &&  0.02  &&  $-0.009$\\
Nuclear structure  & & $\sim 0$ &  & $\sim0$ &&  $-0.002$ \\
\hline 
Final  &&  $2.8(2) \!\times\! 10^{-4}$  &&  $33.9(1)$  &&  $3.14(2)$ \\
\hline 
\end{tabular}
\end{center}
\end{table}

Quantitative light shifts values can be predicted by taking into account geometric factors \cite{Sahoo2016}, and by assigning electric field amplitudes to the standing waves in Eqs.\ (\ref{eq_E2PNCfield}). For a realistic example, we have assumed laser powers at $\lambda_\mathrm{sd}$ of 3 W, coupled into a cavities with enhancement factors of 100, and focused to beam diameters of 0.5 mm. This will result in electric field amplitudes of $\mathcal{E}_\mathrm{E2} \!=\! \mathcal{E}_\mathrm{PNC} \!\approx\! 2 \!\times\! 10^{6} \: \mathrm{V}/\mathrm{m}$, (same as in \cite{Fortson1993}). Resulting Rabi frequencies are presented in Table \ref{tab_Omegas}. 
\begin{table}
\begin{center}
\caption{Calculated energy shifts for relevant $\Delta M\!=\!0$ transitions on the spectral line $6\sr\,^2\SR_{1/2},F\!=\!4$--$5\dr\,^2\DR_{3/2},F\!=\!5$, using the amplitudes from Table \ref{tab_transamp} and electric field amplitudes of $2 \!\times\! 10^{6} \: \mathrm{V}/\mathrm{m}$.}
\label{tab_Omegas}
\ra{1.3}
\begin{tabular}{ccccc}
\hline
&\phantom{abc}&  $W_\mathrm{E2}/h$   &\phantom{ab}&  $W_\mathrm{PNC}/h$  \\ 
\hline
$M_F\!=\!1$--$M_F^\prime\!=\!1$  &&  $-7.41$ MHz  &&  $-0.355$ Hz  \\ 
$M_F\!=\!3$--$M_F^\prime\!=\!3$  &&  $-18.16$ MHz  &&  $-0.290$ Hz  \\ 
$M_F\!=\!4$--$M_F^\prime\!=\!4$  &&  $-18.16$ MHz    &&  $-0.217$ Hz    \\ 
\hline
\end{tabular}
\end{center}
\end{table} 
With parameters as above, the predicted signal is: $\omega_\mathrm{obs}/2\pi\!\approx\!0.9$ Hz. 

For Cs, a number density of $10^{12}$ cm$^{-3}$ is attainable. With E2 and PNC beam diameters as above, and estimating the interaction volume as the cube of that, it is possible to hold $>\!10^8$ atoms. The radiative lifetime of the upper state is 909 ns \cite{Sahoo2016B}. This means that for a 10\% sensitivity in $\omega_\mathrm{obs}$, the minimum total interrogation time is $\approx\!30$ ms. A more ambitious benchmark is a comparison with \cite{Wood1997}. Combining experimental data from \cite{Wood1997} with theory from \cite{Dzuba2012} yields a weak charge for Cs of $Q_\mathrm{w}\!=\!72.58$ with an uncertainty of $0.6\%$. In our case, $Q_\mathrm{w}$ will be proportional to $\omega_\mathrm{obs}$, which, given the same quality of theoretical data as in \cite{Dzuba2012}, means that a corresponding sensitivity would require a resolution better than 5 mHz. An improvement of that to a precision better than 1 mHz, will for the estimated parameters require a total interrogation time of 30000 s. 

The minimum magnetic field necessary in order to tune $\Delta M_F\!=\!2$ E2 transitions out of resonance is $B\!=\!3$ mT. This means a total quadratic Zeeman shift contribution from Eq \. (\ref{eq_quadraticZeeman}) of $2\pi\!\times\!150$ kHz. This can be accurately measured by RF spectroscopy, and with the same technique, and added screening, the field can be stabilised. The required level of magnetic field control is $10^{-6}$. 
A small birefringence in the optical viewports will lead to small circular polarisation components in the light, which will produce small imbalances in the cancellations of optical lattice and E2 light shifts in Eq.\ (\ref{eq_OmegaObs}). However, by measuring the level shifts first with only the optical lattice turned on, and then the optical lattice and the E2 standing wave, these can be corrected for, and techniques for quantifying induced ellipticity \textit{in situ} have been reported in \textit{e.g.\ } \cite{Steffen2013,Park2008}.     
The atoms must be confined to a small volume, where conditions for detection are met. To estimate the achievable localisation, we assume optical lattice lasers at $\lambda_\mathrm{ol}$ with 5 W per beam, collimated to diameters of 1 mm. The dominating contributions to the optical lattice light shift emanate from the D1 and D2 lines, and with above parameters, the potential depth is $U_0 \!\approx\! h \!\times\! 1$ MHz \cite{Grimm2000}. The optical lattice loads from a 3D optical molasses, with which Cs temperatures of 1 $\mu$K (corresponding to 10 kHz) are routinely achieved. This gives a linear confinement in the detection plane of the order of $\lambda_\mathrm{ol}/30$. The shortest separation between two detection points is $\lambda/(2\sqrt{2})$, and a maximally unfavourable point for detection will be at half that distance. 

Our analyses of a predicted signal show that the proposed experimental scheme is a fordable route for detection of atomic PNC and NAM. The computed PNC amplitude in Table \ref{tab_transamp} for the $6\sr\,^2\SR_{1/2}$--$5\dr\,^2\DR_{3/2}$ line is about 3.5 times larger than that for $6\sr\,^2\SR_{1/2}$--$7\sr\,^2\SR_{1/2}$ \cite{Safronova2018}. The specific example with a NSI effect in Cs is used for a feasibility study, but the scheme is not limited to this. The proposal may be used also for NSD experiments, other optical lattice geometries than the one in Fig.\ \ref{fig_setup} are possible, and the general experimental idea is applicable also to other species than Cs. In terms of the potential sensitivity to new physics beyond the SM, for the particular suggested NSI measurement, a limitation is the theoretical uncertainty in the calculated $A_\mathrm{PNC}$ in Table \ref{tab_transamp} of 0.6\%. The accuracy achieved in this calculation is a substantial improvement over the previous calculations \cite{Sahoo2017,Dzuba2001}, and there is scope for further improvements. This means that if a measured value of the PNC light shift, with added experimental uncertainty, would yield a value of $Q_\mathrm{w}$ differing from the predicted value with more than 0.6\%, that would be an indication of new physics. That translates to $\Delta Q_\mathrm{w}\!>\!0.44$, which in turn would mean a particle of mass $>\!3.3$ TeV/c$^2$ \cite{Aoki2017}.

{\it Acknowledgement}: 
Computations were carried out using Vikram-100 HPC of Physical Research Laboratory, India.


%

\end{document}